\documentclass[conference]{IEEEtran}
\IEEEoverridecommandlockouts
\usepackage{amsmath,amssymb,amsfonts}
\usepackage{algorithmic}
\usepackage{graphicx}
\usepackage{textcomp}
\usepackage{xcolor}
\usepackage{multirow}
\def\BibTeX{{\rm B\kern-.05em{\sc i\kern-.025em b}\kern-.08em
    T\kern-.1667em\lower.7ex\hbox{E}\kern-.125emX}}

\begin{document}

\title{Planar Gaussian Splatting with Bilinear Spatial Transformer for Wireless Radiance Field Reconstruction
}
\author{
    \IEEEauthorblockN{Jinghan Zhang$^{\star \ddagger}$, Xitao Gong$^{\star}$, Qi Wang$^{\star}$, Richard A. Stirling-Gallacher$^{\star}$, Giuseppe Caire$^{\ddagger}$}
    \IEEEauthorblockA{
        $^{\star}$Munich Research Center, Huawei Technologies Duesseldorf GmbH, Munich, Germany \\
        $^{\ddagger}$Department of Electrical Engineering and Computer Science, Technical University of Berlin, Berlin, Germany
    }
}

\maketitle
\begin{abstract}
Wireless radiance field (WRF) reconstruction aims to learn a continuous, queryable representation of radio frequency characteristics over 3D space and direction, from which specific quantities—such as the spatial power spectrum (SPS) at a receiver given a transmitter position—can be predicted. While Gaussian splatting (GS)-based method has surpassed Neural Radiance Fields (NeRF)-based method for this task, existing adaptations largely transplant vision pipelines, limiting physical interpretability and accuracy. We introduce BiSplat-WRF, a planar GS framework that retains the expressiveness of 3D GS while removing unnecessary projections and incorporating global EM coupling and mutual scattering among primitives. Each primitive is a 2D planar Gaussian with 3D coordinates, rendered directly on the angular domain of the SPS. A bilinear spatial transformer (BST) aggregates inter-primitive relations on an angular grid and, via attention, captures long-range electromagnetic dependencies, thereby enforcing globally aware EM interactions that reflect the complex physics of the wireless environment. On spatial spectrum synthesis task, BiSplat-WRF surpasses NeRF-based and prior GS-based baselines with respect to the Structural Similarity Index (SSIM); comprehensive ablation studies validate the contribution of BST. We also provide a larger BiSplat-WRF+ variant that further increases SSIM at a higher computation cost, serving as a strong reference for future studies.
\end{abstract}


\section{Introduction}
Modern wireless communication systems require accurate understanding of radio frequency (RF) signal propagation in complex environments for optimal network design and deployment. Traditional approaches, including probabilistic channel models~\cite{Prob} and ray-tracing simulations~\cite{RT, Sionna}, face significant limitations: probabilistic models cannot capture site-specific and material-dependent multipath propagation, while ray-tracing requires precise geometric and material properties which are challenging to obtain in practice.

Recently, learning-based approaches such as neural surrogate ray-tracing~\cite{digital_twin, GeNeRT, reflectance} and Wireless Radiance Fields (WRF) have emerged as promising alternatives. They provide cost-effective pathways for generating large-scale synthetic data while offering detailed channel characteristics through accurate propagation modeling.

NeRF$^2$~\cite{nerf2} pioneered the adaptation of Neural Radiance Fields (NeRF) from computer vision to the RF domain, achieving notable success in WRF reconstruction. However, NeRF-based methods suffer from computational complexity due to volumetric ray sampling. In computer vision, 3D Gaussian Splatting (3D GS)~\cite{3DGS} has increasingly replaced neural radiance fields, offering rendering speed improvements of several orders of magnitude~\cite{3DGS_survey1,3DGS_survey2}. This advantage has been successfully leveraged in wireless communications through WRF reconstruction methods~\cite{WRFGS, RFSPM}. Recent advances including WRF-GS+~\cite{WRFGS+} and SwiftWRF~\cite{SwiftWRF} demonstrate superior performance compared to NeRF$^2$, with WRF-GS+ achieving better accuracy while SwiftWRF provides remarkable efficiency.

While 3D GS-based WRF reconstruction algorithms effectively utilize the expressive power of Gaussian primitives and GPU-accelerated parallel rendering, they typically treat each primitive as an isolated emitter, failing to account for the global electromagnetic (EM) coupling inherent in wireless propagation. In physical environments, the signal contribution of a specific scatterer is not independent; rather, it is the cumulative result of complex interactions, such as secondary reflections and mutual scattering among all objects in the scene. To address this, we introduce BiSplat-WRF, which employs a Bilinear Spatial Transformer (BST) to model these inter-primitive EM dependencies, enabling each Gaussian to become globally aware of the environmental scattering context. Furthermore, our framework avoids the unnecessary computational overhead associated with direct Cartesian coordinates~\cite{WRFGS+,RFSPM}, while overcoming the limitations of purely 2D planar approaches~\cite{SwiftWRF} that lack depth representation and the ability to model essential occlusion relationships.

To further enhance the expressive capability of Gaussian splatting for WRF, this paper presents \textbf{BiSplat-WRF}, a novel neural representation framework that leverages a BST to model global electromagnetic interactions. By facilitating inter-primitive coupling, BiSplat-WRF enables a globally aware formulation that addresses the fundamental limitations of isolated primitive modeling. Our key contributions include:

\begin{itemize}
\item \textbf{Angular-domain Gaussian Representation}: We employ 2D planar Gaussian primitives parameterized directly in 3D azimuth-e
elevation-depth coordinates. This formulation omits unnecessary spatial projections by operating in the angular domain of the SPS, while retaining the depth information required for accurate occlusion modeling and distance-dependent scaling.

\item \textbf{Global Context Modeling}: We propose a BST architecture that enables each Gaussian primitive to capture global inter-primitive relationships, providing physically meaningful representations of electromagnetic wave interactions.

\item \textbf{Multi-Scale Hierarchy-based Mixing}: We split the predictor into static and dynamic sub-networks while organizing Gaussian primitives into coarse- and fine-scale bases, capturing both the stable background propagation patterns and the dynamic multi-path fluctuations inherent in complex wireless environments.

\item \textbf{Comprehensive Evaluation}: We demonstrate superior performance on spatial power spectrum (SPS) synthesis with state-of-the-art Structural Similarity Index (SSIM) scores, showing the effectiveness of BST architecture through ablation studies.
\end{itemize}

The remainder of this paper is organized as follows: Section~\ref{sec:Preliminaries} provides preliminaries on WRF and 3D GS. Section~\ref{sec:Problem} formulates the problem of SPS prediction. Section~\ref{sec:Method} details our BiSplat-WRF methodology. Section~\ref{sec:exp_results} presents comprehensive experimental results on SPS synthesis. Section~\ref{sec:conl} concludes the paper and discusses future research directions.

\begin{figure}[b]
  \centering
  \includegraphics[width=0.5\textwidth]{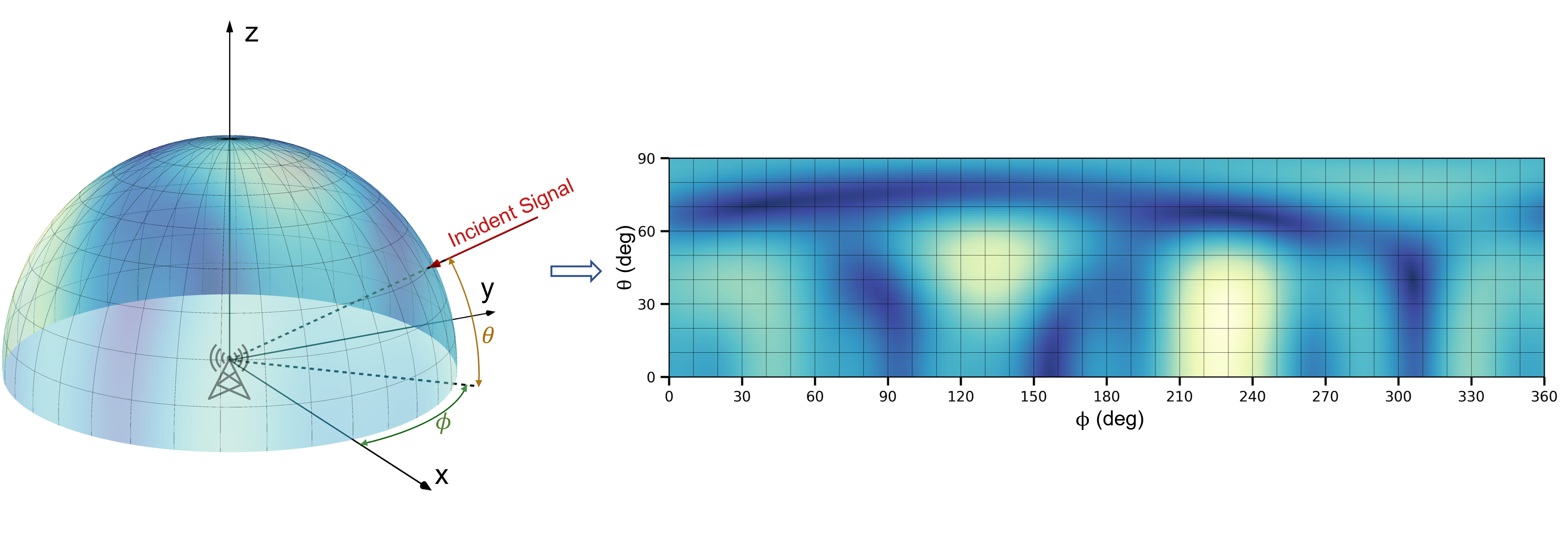}
  \caption{Illustration of spatial power spectrum}
  \label{fig:SPS}
\end{figure}
\begin{figure*}[!t] 
  \centering
  \includegraphics[width=0.9\textwidth]{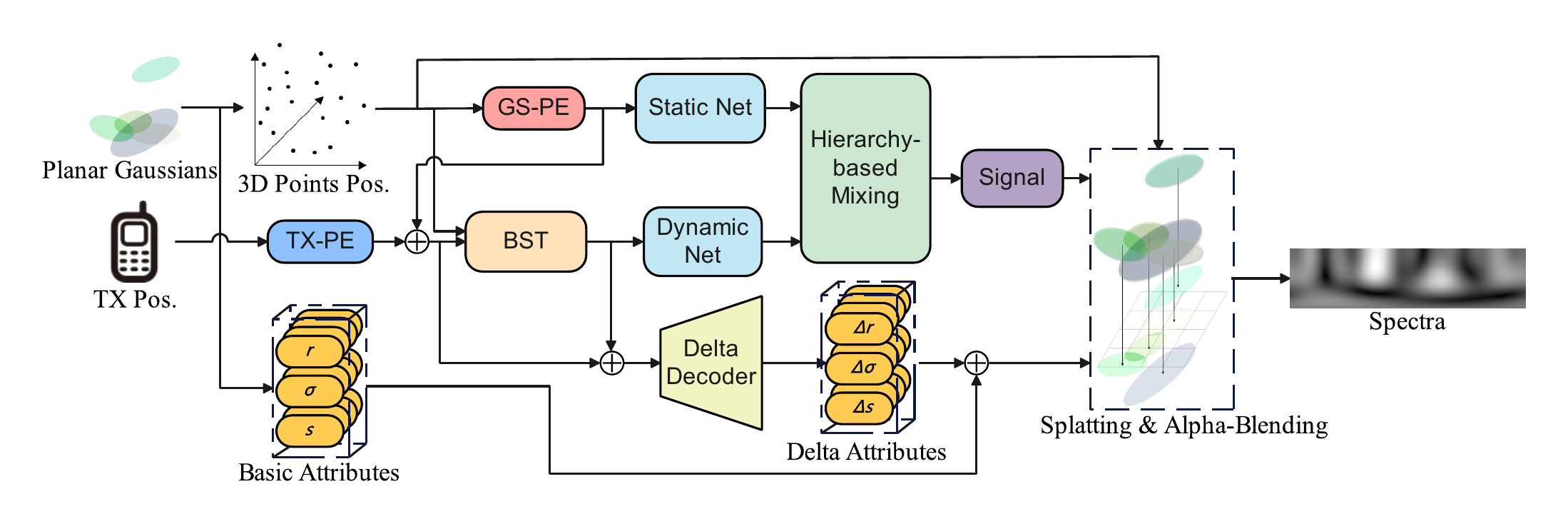} 
\caption{Overall framework of BiSplat-WRF. The architecture integrates static and dynamic signal modeling with a BST for global EM coupling. Signal values are derived through hierarchy-based mixing, and attribute offsets are superimposed on base Gaussian parameters for rendering. The BST module is detailed in Fig.~\ref{fig:bst}.}

  \label{fig:framework}
\end{figure*}

\section{Preliminaries}
\label{sec:Preliminaries}
\subsection{Wireless Radiance Field and Spatial Power Spectrum}
A Wireless Radiance Field (WRF)~\cite{WRFGS, WRFGS+, nerf2, RF_PGS} represents a queryable, continuous field over 3D space and direction that encodes how RF energy is distributed and transported in an environment. Analogous to optical radiance fields, WRF captures the spatial characteristics of wireless signals, enabling synthesis or reconstruction of channel responses and spatial spectrum at arbitrary transmitter–receiver configurations. Unlike visible-light fields, WRFs model radio signals in the MHz–GHz range, whose longer wavelengths induce pronounced multipath effects via reflection, diffraction, and scattering from scene geometry~\cite{6Gsurvey}.

The SPS serves as a fundamental representation of WRF characteristics, characterizing the angular distribution of received signal power across receiving directions. For a receiver equipped with an $M \times N$ uniform planar antenna array, the SPS at direction $(\phi, \theta)$ is computed through beamforming:

\begin{equation}
P(\phi, \theta) = \frac{1}{MN}\left| \sum_{m=0}^{M-1} \sum_{n=0}^{N-1} y_{m,n} e^{-j\Delta\sigma_{m,n}(\phi,\theta)} \right|^2
\end{equation}

where $y_{m,n}$ represents the complex signal received at antenna element $(m,n)$, and $\Delta\sigma_{m,n}(\phi,\theta)$ is the geometric phase shift relative to the reference antenna $(0,0)$.

\subsection{3D Gaussian Splatting}
3D Gaussian Splatting models a scene using a collection of 3D anisotropic Gaussian primitives distributed throughout the space. This explicit representation offers superior computational efficiency compared to implicit neural fields while maintaining high rendering quality. Each individual Gaussian $G(\boldsymbol{x})$ is mathematically defined as:

\begin{equation}
G(\boldsymbol{x}) = e^{-\frac{1}{2}(\boldsymbol{x}-\boldsymbol{\mu})^T\boldsymbol{\Sigma}^{-1}(\boldsymbol{x}-\boldsymbol{\mu})}
\end{equation}

where $\boldsymbol{x} = [x_0, x_1, x_2]^T$ represents the spatial coordinates, $\boldsymbol{\mu} \in \mathbb{R}^3$ is the mean vector, and $\boldsymbol{\Sigma} \in \mathbb{R}^{3 \times 3}$ is the covariance matrix. The covariance matrix is parameterized as $\boldsymbol{\Sigma} = \boldsymbol{R}\boldsymbol{S}\boldsymbol{S}^T\boldsymbol{R}^T$, where $\boldsymbol{R}$ is a rotation matrix represented by quaternions $\boldsymbol{q} \in \mathbb{R}^4$, and $\boldsymbol{S}$ is a diagonal scaling matrix with elements $\boldsymbol{s} \in \mathbb{R}^3$. Each Gaussian also possesses opacity $\alpha \in \mathbb{R}$ and Spherical Harmonics (SH) coefficients $\boldsymbol{\tau} \in \mathbb{R}^{3 \times (k+1)^2}$ (where k is the degree of the SH coefficients) for view-dependent color modeling:

\begin{equation}
\boldsymbol{c} = \text{sh}(\boldsymbol{\tau}, \boldsymbol{v}, k) \in \mathbb{R}^3
\end{equation}

where $\boldsymbol{v} \in \mathbb{R}^3$ is the normalized view direction vector from the Gaussian center to the camera, enabling view-dependent appearance modeling.

During rendering, 3D Gaussians are projected onto the image plane forming 2D splats through rasterization. Each 2D splat $w_i(p)$ represents the projected contribution weight of the $i$-th Gaussian at pixel $p$, computed as a 2D Gaussian distribution:

\begin{equation}
w_i(p) = \exp\left(-\frac{1}{2}(\boldsymbol{p} - \boldsymbol{\mu}'_i)^T{\boldsymbol{\Sigma}'_i}^{-1}(\boldsymbol{p} - \boldsymbol{\mu}'_i)\right)
\end{equation}

where $\boldsymbol{\mu}'_i$ and $\boldsymbol{\Sigma}'_i$ are the projected 2D mean and covariance matrix obtained through the affine transformation $\boldsymbol{\Sigma}'_i = \boldsymbol{J}\boldsymbol{W}\boldsymbol{\Sigma}_i\boldsymbol{W}^T\boldsymbol{J}^T$, with $\boldsymbol{J}$ being the Jacobian matrix and $\boldsymbol{W}$ the view transformation matrix. The final pixel color is computed via alpha-blending, which accumulates contributions from all overlapping Gaussians in depth order:

\begin{equation}
\label{eq:alpha_blending}
\boldsymbol{C}(p) = \sum_{i \in I_p} \boldsymbol{c}_i \cdot \alpha_i  w_i(p) \prod_{j=1}^{i-1}(1 - \alpha_j w_j(p))
\end{equation}

where $I_p$ denotes the set of Gaussians affecting pixel $p$, and the Gaussians are sorted front-to-back for proper depth ordering in the blending process.

\section{Problem Formulation}
\label{sec:Problem}
We consider a wireless scenario where a receiver with a uniform planar array is fixed at a reference location. The transmitter can be placed at arbitrary coordinates $\mathbf{r}_{TX} \in \mathbb{R}^3$. Given $\mathbf{r}_{TX}$, the task is to predict the SPS $\mathbf{P} \in \mathbb{R}^{360 \times 90}$ at the receiver, as in Fig.~\ref{fig:SPS}.

Formally, we learn
\begin{equation}
\label{eq:mapping}
f_\theta: \mathbb{R}^3 \rightarrow \mathbb{R}^{360 \times 90}
\end{equation}
where $f_\theta(\mathbf{r}_{TX}) = \mathbf{P}$ is the predicted spectrum with one-degree resolution over azimuth $\phi \in [0^\circ,360^\circ)$ and elevation $\theta \in [0^\circ,90^\circ]$, following conventional settings~\cite{nerf2,WRFGS+,SwiftWRF}.

The dataset $\mathcal{D}=\{(\mathbf{r}_{TX}^{(k)},\mathbf{P}^{(k)})\}_{k=1}^K$ comprises $K$ TX–spectrum pairs collected in a static propagation environment; training/testing splits are drawn within this same environment. Each entry $P^k(\phi,\theta)$ encodes received power from direction $(\phi,\theta)$ relative to the array, capturing multipath effects. The scope is per-environment: a different environment requires a separately trained (or fine-tuned) model.

Prediction quality is measured by  Structural Similarity Index (SSIM), comparing luminance, contrast, and structure between predicted and ground-truth spectrum. SSIM well reflects perceptual similarity in the spatial distribution patterns that are critical for wireless channel characterization.


\begin{figure*}[!t] 
  \centering
  \includegraphics[width=\textwidth]{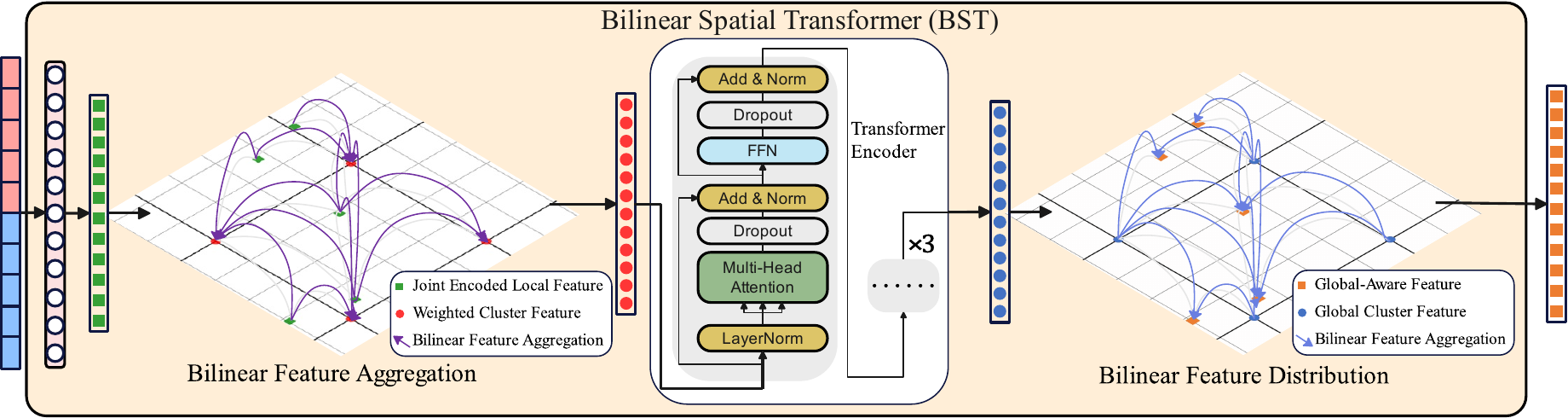} 
  \caption{Illustration of BST. It takes the concatenation of GS-PE and TX-PE as input and projects to joint encoded local feature. A bilinear feature aggregation is then implemented to converge these features to weighted cluster features. A multi-layer multi-head attention captures the global information for each feature. Finally the global cluster feature is mapped to global-aware feature for each Gaussian primitive, which is the output of the BST.}
  \label{fig:bst}
\end{figure*}

\section{Methodology}
\label{sec:Method}
\subsection{Overall Framework of BiSplat-WRF}

In BiSplat-WRF, each primitive $i$ is modeled as a 2D Gaussian distribution parallel to the $(\phi, \theta)$ plane but augmented with a depth coordinate, yielding a planar representation in the angular domain that supports direct tile-based rasterization while preserving depth for occlusion modeling. The $i$-th Gaussian primitive has parameters $\boldsymbol{x}_{g,i} = [\phi_i, \theta_i, d_i]^{\top}$, where $\phi_i \in [0, 360)$ and $\theta_i \in [0, 90]$ denote the azimuth and elevation angles, respectively, and $d_i > 0$ represents a monotonic depth proxy (increasing with range); in our implementation it represents the scatterer–RX distance and is used for depth ordering and distance-dependent scaling. Its shape is controlled by the scale factors $\boldsymbol{s}_i = [s_{x,i}, s_{y,i}]^{\top}$ with $s_{x,i}, s_{y,i} > 0$, together with an in-plane rotation angle $\rho_i \in [0, \pi)$. Each primitive also carries an opacity $o_i \in (0, 1)$ to model the attenuation of electromagnetic waves by intervening objects. All these basic attributes are learnable and optimized during the training process.

The projection of planar primitive $i$ onto the $(\phi, \theta)$ plane is characterized by:
\begin{equation}
G_i(\phi, \theta) =
\exp\!\left(
-\tfrac{1}{2}(\boldsymbol{p} - \boldsymbol{\mu}_i)^{\top}
\boldsymbol{\Sigma}_i^{-1}
(\boldsymbol{p} - \boldsymbol{\mu}_i)
\right),
\end{equation}
where $\boldsymbol{p} = [\phi, \theta]^{\top}$ and $\boldsymbol{\mu}_i = [\phi_i, \theta_i]^{\top}$; the covariance $\boldsymbol{\Sigma}_i$ is derived from the previously defined $\boldsymbol{s}_i$ and $\rho_i$.

Unlike traditional 3D GS where signals are directly associated with individual primitives, our approach generates the complex signal coefficient $c_i = a_i + j b_i \in \mathbb{C}$ through dedicated neural networks that consider global scene context and transmitter-dependent interactions. This decoupling enables physically meaningful modeling of electromagnetic wave propagation, where signals result from complex interactions between multiple scatterers rather than being isolated to individual Gaussian primitives.

As illustrated in Fig.~\ref{fig:framework}, BiSplat-WRF transforms the TX position into the spatial power spectrum using positional embeddings for primitive and TX coordinates (GS-PE and TX-PE). A \emph{Static Network} takes GS-PE as input to produce TX-independent signal components, while the BST processes the concatenation of GS-PE and TX-PE to model global EM coupling. The global-aware features output by the BST then serve as the input to the \emph{Dynamic Network} for predicting TX-dependent signal modulations. Simultaneously, a \emph{Delta Decoder} predicts attribute offsets (rotation, scale, and opacity) by leveraging both the concatenated embeddings and the BST's global features; this design ensures that local attribute adjustments are informed by both the specific TX-primitive geometry and the broader environmental scattering context. These modules all utilize a residual feed-forward network (FFN) structure. The outputs from the \emph{Static Network} and \emph{Dynamic Network} are combined via Hierarchy-based Mixing to obtain the signal value represented by each Gaussian primitive, while the rotation, scale, and opacity offsets for each Gaussian primitive are superimposed onto the base values of these attributes.

Rendering proceeds directly on the $(\phi,\theta)$ plane with depth ordering. Alpha-blending is then applied as Eq.~\ref{eq:alpha_blending}, separately accumulating the real and imaginary parts of the complex field. The SPS at the RX is obtained per pixel as the magnitude squared of the accumulated field.

\subsection{BST for Global Context Modeling}
\begin{figure}[tp]
  \centering
  \includegraphics[width=0.3\textwidth]{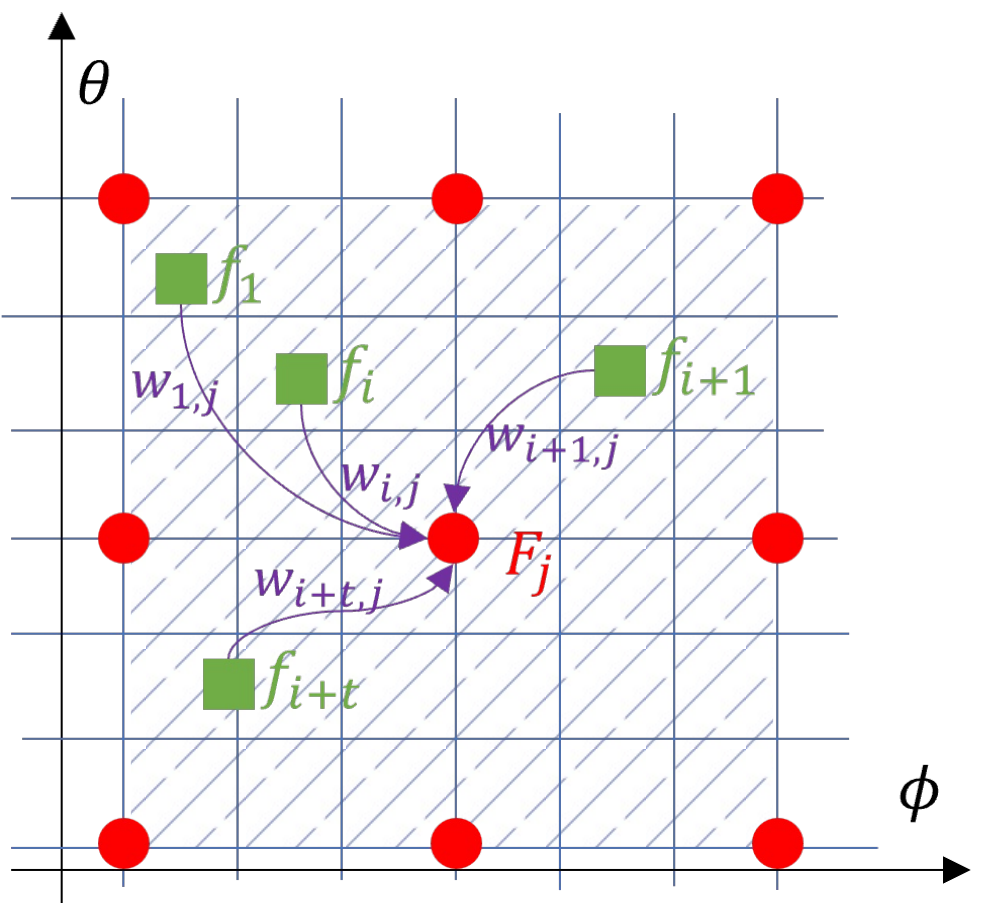} 
  \caption{Illustration of bilinear feature aggregation in BST, where the region with dashed background is the coverage area of a grid point $j$.}
  \label{fig:bst-area}
\end{figure}
To bridge the gap between isolated primitives and the physical reality of multi-path coupling, we introduce the BST module to explicitly model global EM interactions. This architecture enables each primitive to account for the collective influence of other primitives, facilitating the representation of secondary reflections and mutual scattering. The overall process of BST is shown in Fig.\ref{fig:bst}. BST first transforms concatenated GS-PE and TX-PE into joint encoded local features $\boldsymbol{F}=[\boldsymbol{f}_1, \boldsymbol{f}_2, \ldots, \boldsymbol{f}_{N_p}]^T$ (green squares in Fig.~\ref{fig:bst}) of total $N_p$ primitives through a linear projector. Then $\boldsymbol{f}_i$s are assigned to 4 nearest spatial cluster grid. As shown in Fig~\ref{fig:bst-area} the angular domain is partitioned into $N_c$ clusters arranged in a $\sqrt{N_c} \times \sqrt{N_c}$ grid with spacing $\Delta\phi = \frac{360^\circ}{\sqrt{N_c}}$ and $\Delta\theta = \frac{90^\circ}{\sqrt{N_c}}$. For each primitive $i$ at position $(\phi_i, \theta_i)$ and grid point $j$ at $(\phi_j, \theta_j)$, the bilinear interpolation weight $w_{i,j}$ is computed by accounting for the periodicity of the azimuth angle. We first define the angular distance for the azimuth as $d_\phi(i, j) = \min(|\phi_i - \phi_j|, 360^\circ - |\phi_i - \phi_j|)$. The weight is then expressed as:
\begin{equation}
w_{i,j} = \max\left(0, 1 - \frac{d_\phi(i, j)}{\Delta\phi} \right) \cdot \max\left(0, 1 - \frac{|\theta_i - \theta_j|}{\Delta\theta} \right)
\end{equation}

The set of primitives associated with grid point $j$ is shown with dashed area in Fig\ref{fig:bst-area} and is defined as:

\begin{equation}
\mathcal{S}_j = \{i : |\phi_i - \phi_j| \leq \Delta\phi \text{ and } |\theta_i - \theta_j| \leq \Delta\theta\}
\end{equation}

The weighted cluster feature (Red dots in Fig.~\ref{fig:bst}) $\boldsymbol{g}_j$, which contains the information from the primitives within $\mathcal{S}_j$ is then computed as the normalized weighted sum:

\begin{equation}
\boldsymbol{g}_j = \frac{\sum_{i \in \mathcal{S}_j} w_{i,j} \boldsymbol{f}_i}{\sum_{i \in \mathcal{S}_j} w_{i,j}}
\end{equation}

These weighted cluster features $\boldsymbol{G}=[\boldsymbol{g}_1, \boldsymbol{g}_2, \ldots, \boldsymbol{g}_{N_c}]^T$ are processed through a multi-head Transformer encoder~\cite{vaswani2017attention} to produce Global Cluster Features $\boldsymbol{G}'=[\boldsymbol{g'}_1, \boldsymbol{g'}_2, \ldots, \boldsymbol{g'}_{N_c}]$ (Blue dots in Fig.~\ref{fig:bst}) that capture scene-wide electromagnetic interactions. This Transformer has multiple layers that have the same structure. Finally, a bilinear feature distribution process shown on the right side of Fig.~\ref{fig:bst} maps each $\boldsymbol{g}_i'$ back to global-aware Feature $\boldsymbol{f}_i'$ of individual primitive $i$ within $\mathcal{S}_j$ using weights $w_{ij}$ through an inverse bilinear interpolation:

\begin{equation}
\boldsymbol{f}'_i = \frac{\sum_{j| i \in \mathcal{S}_j} w_{i,j} \boldsymbol{g}'_j}{\sum_{j| i \in \mathcal{S}_j} w_{i,j}}
\end{equation}

$\boldsymbol{F}'=[\boldsymbol{f}_1', \boldsymbol{f}_2', \ldots, \boldsymbol{f}_{N_p}']^T$ has exactly the same dimension as $\boldsymbol{F}$ but encodes both local characteristics and global context for subsequent \emph{Dynamic Network} and \emph{Delta Decoder}.

The steps of the bilinear feature aggregation and distribution are important here to keep the complexity of the input dimension of the Transformer under control. It keeps the Transformer input length fixed at $N_c$ (independent of $N_p$), reducing attention complexity from $\mathcal{O}(N_p^2)$ to $\mathcal{O}(N_c^2)$; with a fixed grid, this is constant per query.

\subsection{Multi-Scale Hierarchy-based Mixing}
Our framework decomposes the signal representation into static and dynamic components to differentiate between static environmental geometry and TX-dependent propagation effects. This separation allows the model to simultaneously capture stable background propagation patterns and dynamic multi-path fluctuations. To align these components with the physical scale of the environment, we organize Gaussian primitives into coarse-scale and fine-scale groups according to their size. The complex signal coefficient $c_i$ for each primitive is fused as follows:
\begin{equation}
c_i = (\text{amp}_{\text{static},i} + w_{\text{scale},i} \cdot \text{amp}_{\text{mod},i}) \cdot e^{j(\text{phase}_{\text{static},i} + \text{phase}_{\text{mod},i})}
\end{equation}
where the $\text{amp}$ and $\text{phase}$ terms are real scalars. The weighting factor $w_{\text{scale},i}$ acts as a scale-dependent coupling mechanism: for coarse-scale primitives modeling large structural surfaces, $w_{\text{scale},i}$ is assigned a smaller value to prioritize the static component, ensuring the stability of background propagation. Conversely, for fine-scale primitives modeling microscopic details, $w_{\text{scale},i}$ is increased to amplify the dynamic modulation, thereby capturing the high sensitivity of local scattering to the transmitter's movement. This multi-scale fusion strategy ensures a physically consistent reconstruction of the wireless radiance field across different spatial resolutions.

\section{Implementation And Evaluation}
\subsection{Implementation Details}
\textbf{Initialization}: 
Gaussian primitives are randomly distributed in the angular-depth space with azimuth and elevation angles uniformly sampled as $\phi \sim \mathcal{U}(0^\circ, 360^\circ)$ and $\theta \sim \mathcal{U}(0^\circ, 90^\circ)$. To capture multi-scale electromagnetic interactions, we employ a stratified depth initialization strategy where 30\% of primitives serve as coarse primitives with depths randomly sampled from $d \sim \mathcal{U}(20, 50)$ to model far-field propagation effects, while the remaining 70\% are fine primitives with depths $d \sim \mathcal{U}(1, 20)$ to capture near-field interactions and local multipath phenomena. The initial scaling parameters follow a depth-dependent formulation: $z_{\text{norm}} = (d - 0.1)/(100.0 - 0.1)$ and $s_{\text{base}} = s_{\text{min}} + (s_{\text{max}} - s_{\text{min}}) \cdot \exp(-z_{\text{norm}})$, with scaling axes $(s_x, s_y)$ randomly initialized between 1.0 and 1.5 times $s_{\text{base}}$ to introduce anisotropy. Rotation angles are uniformly sampled from $\mathcal{U}(0, \pi)$, opacity values are set to 0.3.

\textbf{Positional Encoding}: Given an input 3D coordinate vector $\boldsymbol{t} \in \mathbb{R}^3$, the positional embedding after positional encoding is defined as:
\begin{equation}
\gamma(\boldsymbol{t}) = [ \boldsymbol{t}, \{\sin(2^k\pi\boldsymbol{t})\}_{k=0}^{L-1}, \{\cos(2^k\pi\boldsymbol{t})\}_{k=0}^{L-1} ].
\end{equation}
Here, each sine and cosine function is applied element-wise to all dimensions of $\boldsymbol{t}$, and $L$ is the order
of the position encoding. For Gaussian primitive coordinates, we set $L = 8$ since their positions directly influence the reconstruction results and require fine-grained spatial representation. For TX coordinates, we use $L = 6$ since they are inherently imprecise. This yields embedding dimensions of $3\times(2 \times 8 + 1) = 51$ for Gaussian primitives and $3\times(2 \times 6 + 1) = 39$ for TX coordinates.

\textbf{Model Configuration}:
In BiSplat-WRF, all FFNs are implemented as 4-layer residual FFNs with a hidden width of 128. The BST uses 64 clustering grid points, and the Transformer encoder comprises 3 layers with 4 attention heads. We employ a total of $N_p=500$ Gaussian primitives. To probe the capacity limits of our design, we introduce BiSplat-WRF+: all FFNs are expanded to 12 layers with width 256, the Transformer is deepened to 6 layers with internal FFNs of width 512, and the number of Gaussian primitives is increased to 2,000. The overall framework is trained minimizing a composite loss function $\mathcal{L} = 0.7\mathcal{L}_1 + 0.3\mathcal{L}_{\text{SSIM}}$, balancing absolute error and structural similarity.

\subsection{Dataset}
We evaluate our method on the open-source NeRF$^2$ RFID dataset~\cite{nerf2}, widely used in prior wireless radiance field reconstruction~\cite{WRFGS+,SwiftWRF}. It was collected in a laboratory with a USRP-based RX carrying a $4\times4$ antenna array at 915MHz; the transmitter is an RFID tag repeatedly sending RN16 messages. The spatial spectrum is represented as a $360\times90$ matrix over the upper hemisphere with $1^\circ$ angular resolution. We adopt the standard split of 4{,}800 samples for training and 1{,}200 for testing.

\begin{figure}[tp]
  \centering
  \includegraphics[width=0.5\textwidth]{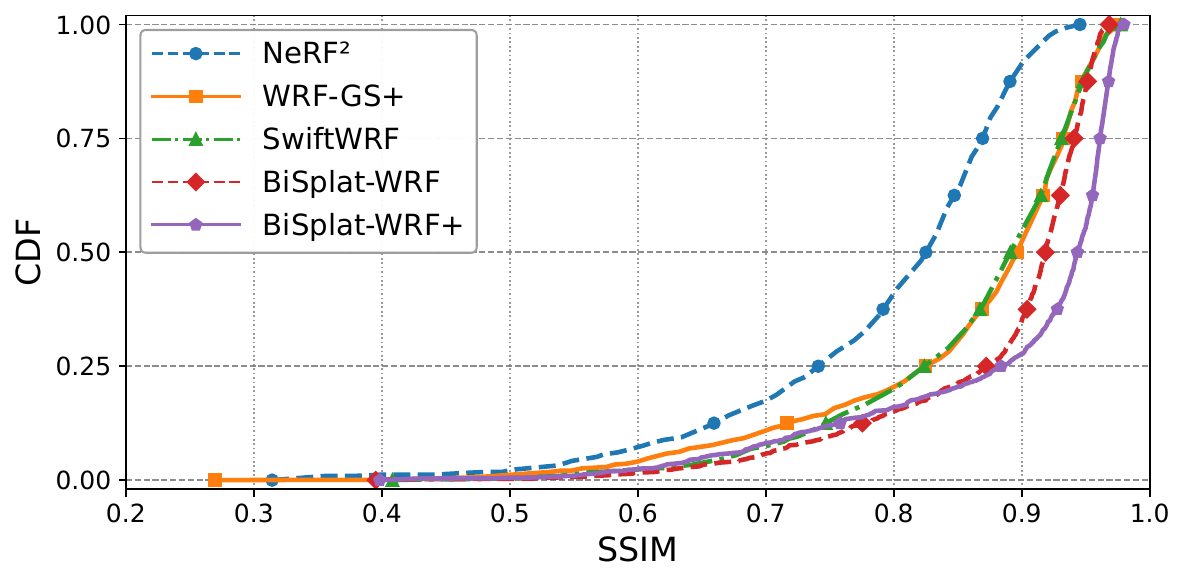} 

  \caption{Cumulative Distribution Function (CDF) of SSIM for various SPS synthesis methods. Our method outperforms all compared baselines by at least 4.4\% in terms of median SSIM.}
  \label{fig:cdf}
\end{figure}

\begin{figure}[tp]
  \centering
  \includegraphics[width=0.5\textwidth]{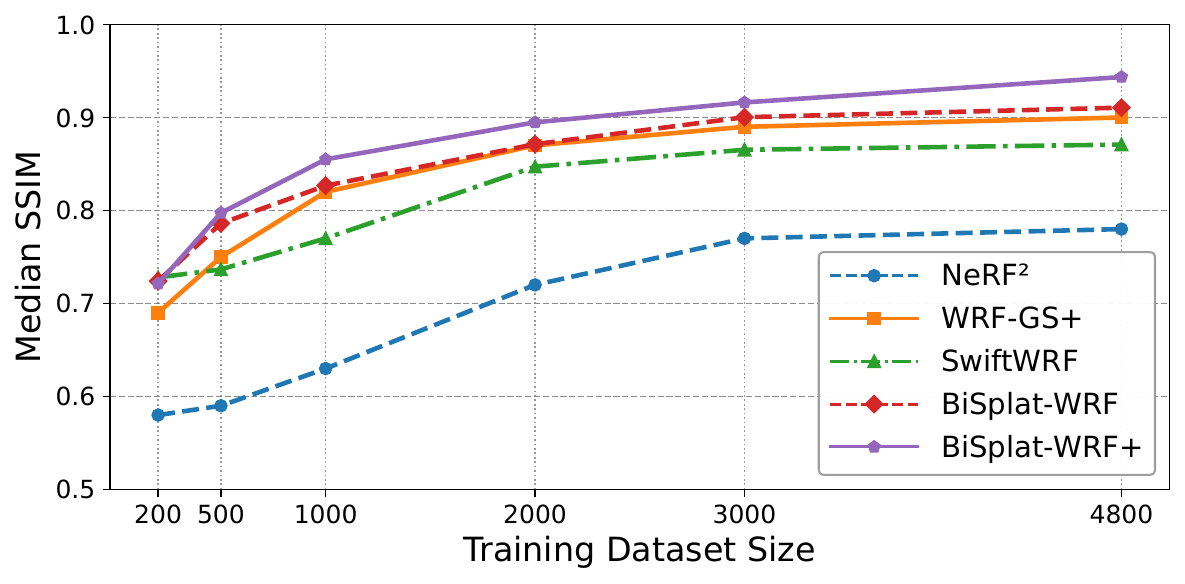} 

  \caption{The median SSIM values of the methods with different training data size and same testing set.}
  \label{fig:datasize}
\end{figure}

\subsection{Experiment Results}
\label{sec:exp_results}
\noindent\textbf{Overall accuracy}:
We compared the results of different existing methods for synthesizing spatial power spectrum.We omit baselines like Ray Tracing, VAE, and DCGAN as they are consistently outperformed by NeRF and GS-based approaches in prior work~\cite{nerf2,WRFGS+,SwiftWRF}. As shown in Fig.~\ref{fig:cdf}, our BiSplat-WRF achieves a median SSIM of 0.91. This represents a clear improvement over NeRF$^2$ (0.82), WRF-GS+ (0.90) and SwiftWRF (0.87). Moreover, the enlarged version, BiSplat-WRF+, further achieves over 4\% improvement over WRF-GS+, reaching a median SSIM of nearly 0.94. We also note that the recently released VoxelRF~\cite{VoxelRF} only reports a median SSIM of 0.90.

\noindent\textbf{Data efficiency}:
Fig.~\ref{fig:datasize} shows performance versus training dataset sizes. Across all training set size ranging from 200 to 4,800 samples and constant testing dataset, BiSplat-WRF maintains a clear SSIM margin over existing approaches. Notably, with about 40\% less data (3,000 samples), BiSplat-WRF+ matches or exceeds WRF-GS+ trained on 4,800 samples, indicating stronger efficiency in data usage and consequently less effort in collecting training data.

\noindent\textbf{Ablation study}: We further evaluate three bypass settings by zeroing the outputs of the corresponding branches (for \emph{Dynamic Network} and \emph{Delta Decoder}) or by directly using the linear projection as the output of BST; results are summarized in Table~\ref{tab:ablation}. When bypassing the \emph{Delta Decoder}, both \textit{BiSplat–WRF} and \textit{BiSplat–WRF+} exhibit similarly small SSIM drops (0.9106$\rightarrow$0.9034 and 0.9435$\rightarrow$0.9366), indicating that the \emph{Delta Decoder} in \textit{BiSplat–WRF} already captures rotation, scale, and opacity offsets sufficiently well, so enlarging it in \textit{BiSplat–WRF+} yields limited additional benefit. Bypassing the \emph{Dynamic Network} brings the two models to comparable performance (0.9066 vs.\ 0.9027), which highlights that the signal-modulation pathway governed by the \emph{Dynamic Network} is essential for pushing the model toward its performance ceiling. In contrast, bypassing the BST causes a pronounced degradation (0.9106$\rightarrow$0.8820 and 0.9435$\rightarrow$0.9045) even though TX positional cues are still provided to other branches, demonstrating that jointly modeling global interactions among Gaussian primitives via BST is crucial for WRF reconstruction and constitutes a core contribution of our approach.
\begin{table}[t]
\centering
\small
\caption{Effectiveness of each module (Median SSIM)}
\label{tab:ablation}
\begin{tabular}{lcc}
\hline
\multirow{2}{*}{Variant} & \multicolumn{2}{c}{SSIM$_{\text{med}}$} \\
\cline{2-3}
& BiSplat\textendash WRF & BiSplat\textendash WRF+ \\
\hline
Full model & \textbf{0.9106} & \textbf{0.9435} \\
Bypass \textit{Dynamic Network} & 0.9066 & 0.9027 \\
Bypass \textit{Delta Decoder} & 0.9034 & 0.9366 \\
Bypass \textit{BST} & 0.8820 & 0.9045 \\
\hline
\end{tabular}
\end{table}

\section{Conclusion}
\label{sec:conl}
We introduce, for the first time in WRF reconstruction, bilinearly interpolated interactions among Gaussian primitives, enabling the global attention module to run in constant time per query. The proposed BiSplat-WRF achieves state-of-the-art performance with a median SSIM of 0.91 for SPS reconstruction. Through ablation studies, we verify that our BST is the key driver of these gains. The scaled variant, BiSplat-WRF+, trades modest increases in computational time and memory for a higher median SSIM of 0.94, which can serve as a performance baseline for future studies. For future work, we will evaluate BiSplat-WRF on applications such as AoA estimation, CSI prediction, and synthetic data generation, and extend validation to additional public datasets while further refining the architecture’s adaptability and generalization.
\bibliographystyle{IEEEbib}
\bibliography{refs}

\end{document}